\documentclass[%
 amsmath,amssymb,
 aps,
pra,
 longbibliography,
 lengthcheck,%
]{revtex4-1}

\usepackage{graphicx}
\usepackage{dcolumn}
\usepackage{color}
\usepackage{bm}
\usepackage{hyperref}

\begin{document}

\preprint{APS/123-QED}

\title{Adhesion of Multi-component Vesicle Membranes}

\author{Yanxiang Zhao}
\email{zhao@math.psu.edu}
\affiliation{Department of Mathematics, Pennsylvania State University, University Park, Pennsylvania, 16802, USA}

\author{Sovan Das}\email{sovandas@iitk.ac.in}
\affiliation{Department of Mechanical Engineering, Indian Institute of Technology, Kanpur 208016, India}

\author{Qiang Du}
\email{qdu@math.psu.edu}
\affiliation{Department of Mathematics and Department of Materials Science and Engineering,\\
   Pennsylvania State University, University Park, Pennsylvania, 16802, USA}

\date{\today}

\begin{abstract}
In this work, we study the adhesion of multi-component vesicle membrane to both flat and curved substrates, based on the conventional Helfrich bending energy for multi-component
vesicles and adhesion potentials of different forms. A phase field formulation is used to describe the different components of the vesicle. For the axisymmetric case, a number of representative equilibrium vesicle shapes are computed and some energy diagrams are
presented which reveal the dependence of the calculated shapes and solution branches on various parameters including both bending moduli and spontaneous curvatures
as well as the adhesion potential constants. Our computation also confirms a recent experimental observation that the adhesion effect may promote phase separation in
two-component vesicle membranes.
\end{abstract}

\pacs{Valid PACS appear here}
\maketitle



\section{\label{sec:level1}Introduction}

Adhesion is a fundamental step for many biological processes such as
exocytosis, endocytosis. Cell adhesion also plays important roles in drug designs
and drug deliveries as well as many biosensor applications  \cite{Dunehoo,lipowsky_book}.
There have been many experimental and theoretical studies focusing on this subject \cite{Das1,Gordon,Lipowsky1,Lipowsky2,Fst07}. While many of the past studies on the vesicle-substrate adhesion have focused on the case of a flat substrate
\cite{seifert1990, seifert1991, seifert1994, seifert1997, smith2003, cantat2003,
 ni2005}, there have also been some works that address the complexity of curved
substrate. For instance, theoretical and experimental studies on the binding of a vesicle membrane to micro or nano-particles, or colloids have been conducted in \cite{dietrich1997, lipowsky1998, deserno2002, deserno2003, deserno2004}, where the characteristic spherical substrates have radii much smaller than that of the vesicles.
In \cite{Das1}, the adhesion of a three dimensional vesicle to curved substrates has been studied where the curvature of the substrates are comparable to the curvature of the vesicles. A phase diagram for bound-unbound transitions has been presented.
In this work, we study the adhesion of multi-component vesicle membranes
to both flat and curved substrates. This is motivated by experimental studies of the modeled subjects. For instance in a recent experiment conducted by Gordon et al.\ \cite{Gordon}, it was observed that a mixed-lipid membrane can go through a local phase separation above critical demixing temperature due to its close proximity to a biological or non-biological surface. That is, adhesion can promote the phase separation for the mixed-lipid cell or vesicle membranes.

In this work, we develop a phase field model to study the adhesion
of multi-component vesicle membranes with a substrate through a
specified adhesion potential. Following our recent approach
described in \cite{JZhang}, we take the adhesion potential to be a
function of distance between the membrane and the substrate. The strength of
adhesion potential is considered to be distinct for different
components. By minimizing the total energy of the system that
includes bending energy, interfacial line tension and the adhesion
energy, the equilibrium vesicle shapes can be computed for a variety
of parameter values. As the initial attempt, we consider the case
that both the vesicle membrane and the substrate are axisymmetric to
simplify the computation. We present, in particular, a number of
typical equilibrium two-component axisymmetric vesicle profiles
undergoing adhesion.  The consistency between the phase field
description and its sharp interface limit is also briefly discussed.
Moreover, a numerical experiment is conducted to support the
conclusion of \cite{Gordon} that the adhesion may promote phase
separation for a multi-component membrane.


\section{Multi-component vesicle membrane with adhesion}
Equilibrium shapes of a multi-component vesicle are often described
by minimizing an energy that includes elastic bending energy of the
membrane and the line tension energy at the interface between the components
\cite{Julicher}. For the elastic bending energy of vesicle membrane, a common form
adopted in the literature is that introduced by Helfrich \cite{Helfrich}:
\begin{align}
E_b=\int_{\Gamma} \Big( \lambda(H-a)^2+b K \Big)\ \text{d}\textbf{x}\,,
\label{helfrich}
\end{align}
where $\Gamma$ is the membrane surface, $H$ and $K$ are the mean and
Gaussian curvatures  of $\Gamma$ with $\lambda$ and $b$ being the mean
curvature bending modulus and the Gaussian curvature bending modulus
respectively and $a$ is the spontaneous curvature. For simplicity,
we consider the effect of mean curvature bending modulus and spontaneous
curvature only. Thus $b$ is set to be zero and Eq.\ \ref{helfrich} becomes
$$
E_b=\int_{\Gamma} \lambda (H-a)^2 \text{d}\textbf{x}\,.
$$

In this paper, we focus on two-component vesicle membranes which
have the liquid-ordered and the liquid-disordered phases, and the
two phases have distinct bending moduli and distinct spontaneous
curvatures \cite{Baumgart1, Baumgart2, Das3}. Let  $\Gamma_1$ and
$\Gamma_2$ be  the parts of the surface representing the two
different phases with $\lambda_1, \lambda_2$ being their
corresponding bending moduli and $a_1, a_2$ being their
corresponding spontaneous curvatures, respectively. The total
elastic bending energy for the two-component vesicle is
\begin{align}\label{Bending}
E_b=\int_{\Gamma_1} \lambda_1 (H-a_1)^2\ \text{d}\textbf{x}+
\int_{\Gamma_2} \lambda_2 (H-a_2)^2\ \text{d}\textbf{x}\,.
\end{align}
The line tension energy, which is essentially an interfacial energy
between the two phases is given as
\begin{align}\label{LineTension}
E_l=\int_{\Gamma_1\cap\Gamma_2}\sigma\ \text{d}l\,,
\end{align}
where $\sigma$ is the constant line tension at the interface.
Eqs (\ref{Bending}) and (\ref{LineTension}) together define the total energy,
\begin{align}\label{FreeTotal}
E=E_b+E_l\,,
\end{align}
with a minimum of $E$ describing the shape of an equilibrium
two-component closed membrane.

Note that the vesicles or membranes discussed so far are free and not
bounded to other objects. To incorporate the adhesive interaction
with a substrate, an additional energetic contribution due to
adhesion should be added to
Eq.\ (\ref{FreeTotal}):
\begin{align}\label{TotalEnergy}
\nonumber
E_{\text{total}}=\int_{\Gamma_1} \lambda_1 (H-&a_1)^2\ \text{d}\textbf{x}+
\int_{\Gamma_2} \lambda_2 (H-a_2)^2\ \text{d}\textbf{x}\\
&+\int_{\Gamma_1\cap\Gamma_2}\sigma\ \text{d}l-\int_{\Gamma} W(\textbf{x})\ \text{d}\textbf{x}\,,
\end{align}
where
\begin{equation}\label{AdhesionPotentialGeneral}
W(\textbf{x})=
 \left\{\begin{array}{l} w_1\cdot P(\textbf{x}),\quad
  \textbf{x}\in\Gamma_1 \\ w_2\cdot P(\textbf{x}),\quad
   \textbf{x}\in\Gamma_2 \end{array}\right.
\end{equation}
 is the adhesion potential which varies with respect to the position
 $\textbf{x}$ on $\Gamma=\Gamma_1\cup\Gamma_2$.
 In the above, $w_1$ and $w_2$ are the corresponding strengths of
the adhesion potential experienced by the liquid-ordered and the
liquid-disordered phases. A representative form of $W$ is that of a
Gaussian form given by,
\begin{equation}\label{AdhesionPotential}
W(\textbf{x})=
 \left\{\begin{array}{l} w_1\exp(-\text{d}(\textbf{x})^2/\epsilon^2),\quad
  \textbf{x}\in\Gamma_1 \\ w_2\exp(-\text{d}(\textbf{x})^2/\epsilon^2),\quad
   \textbf{x}\in\Gamma_2 \end{array}\right.
\end{equation}
where $\text{d}(\textbf{x})$ is the distance from $\textbf{x}$ to a
flat/curved substrate, and $\epsilon$ is a small number. Notice that
when $\epsilon$ approaches zero, the adhesion potential converges to
a sharp contact potential, a scenario that has been investigated in
earlier studies \cite{Das1, Lipowsky2}. While we use the Gaussian
potential (\ref{AdhesionPotential}) in most of this paper, to offer
a comparison, we also consider the Leonard-Jones type potential,
\begin{equation}\label{AdhesionPotentialLJ}
W(\textbf{x})=
 \left\{\begin{array}{l} -w_1\cdot 4\bigg[\bigg(\dfrac{\beta}{\text{d}(\textbf{x})}\bigg)^{\alpha}
-\bigg(\dfrac{\beta}{\text{d}(\textbf{x})}\bigg)^{\alpha/2}\bigg],\  \textbf{x}\in\Gamma_1 \\ -w_2\cdot4\bigg[\bigg(\dfrac{\beta}{\text{d}(\textbf{x})}\bigg)^{\alpha}
-\bigg(\dfrac{\beta}{\text{d}(\textbf{x})}\bigg)^{\alpha/2}\bigg],\  \textbf{x}\in\Gamma_2 \end{array}\right.
\end{equation}
which induces a narrow repulsive region between vesicles and the
substrate. The constant $\beta$ and the exponent $\alpha$ determine
the thickness of the repulsive region and the rate of change of the
adhesion potential, respectively.

\subsection{Phase field formulation}
To be able to effectively describe the different phases of the two-component
vesicle, we use  a phase field formulation which has become very popular
in recent years in the modeling  and simulations of vesicle deformations
\cite{DLW04,DLRW05,misbah,campelo,lowengrub,gao2009,JZhang}.
A phase field function can be used to describe the vesicle with the
phase field bending energy as formulated in \cite{DLW04,DLRW05}. Adhesion
energy can be incorporated into the phase field formulation as shown
in \cite{JZhang}. For multi-component vesicles,
order parameters can be used to describe both the vesicle
and its two components \cite{XWang}. On the other hand,
for a vesicle with a fixed topology, one can also use a direct
(explicit) surface representation for the vesicle along with an
order parameter (phase field function)
to describe the two different phases of the membrane\cite{lowengrub,lowengrub02}. For
the axisymmetric
case considered here, it is particularly effective to adopt a sharp
interface representation of the vesicle surface given by the revolution
of a simple one-dimensional curve with an arc-length parametrization
and a phase field representation of the different phases on the vesicle
which is also a function of the arc-length.

Specifically,  let $\Gamma$ be the vesicle surface, a phase field
function $\eta=\eta(\textbf{x})$ is introduced over $\Gamma$ which
may be used to represent either a material composition profile or a
fictitious density of the lipids on the surface of the membrane and
distinguishes between the liquid-ordered and liquid-disordered
phases. As an illustration, we focus on the latter case so that in
the liquid-ordered phase, $\eta$ is specified to be $+1$ and is
colored as blue in figure \ref{free}; in the liquid-disordered
phase, $\eta$ is assigned to be $-1$ and colored as red. In the
interface between the liquid-ordered and liquid-disordered phases,
$\eta$ rapidly, but continuously, changes from +1 to -1. Note that
the phase field regularizes the sharp interface between the two
different phases into a diffused one, and thus provides a more
general depiction of the two-component vesicle in both the mixed and
de-mixed states. The total energy (\ref{TotalEnergy}) of the model
in terms of the phase function $\eta$ is given by
\begin{align}\label{TotalEnergyPhaseField}
\nonumber
E(\eta)=\int_{\Gamma}(c_0+c_1\eta)&\big[H-a(c_0+c_3\eta)\big]^2\ \text{d}\textbf{x}\\
\nonumber
+\sigma\int_{\Gamma}&\bigg[\dfrac{\xi}{2}\big|\nabla_{\Gamma}\eta\big|^2
+\Phi(\eta)\bigg]\ \text{d}\textbf{x}\\
-&\int_{\Gamma}w(c_0+c_2\eta)
P(\text{d}(\textbf{x}))\ \text{d}\textbf{x}\,,
\end{align}
where the first and third terms are phase field formulas for elastic
bending energy and adhesion energy, respectively. The term
$c_0+c_1\eta$ is a phase field representation of $\lambda_1$ and
$\lambda_2$. When $\textbf{x}$ is away from the interfacial region,
$c_0+c_1=\lambda_1$ and $c_0-c_1=\lambda_2$. The liquid-ordered
phase is stiffer than the liquid-disordered phase, hence we always
assume $c_1>0$. The term $a(c_0+c_3\eta)$ is considered as the phase
field analog of spontaneous curvature, and
\[
a(c_0+c_3)=a_1,\quad  a(c_0-c_3)=a_2
\]
if $\textbf{x}$ is away from the interfacial region. Similarly, $w(c_0+c_2\eta)P(\text{d}(\textbf{x}))$ is viewed as an approximation of
the adhesion potential $W(\textbf{x})$, with
\[
w(c_0+c_2)=w_1,\quad  w(c_0-c_2)=w_2
\]
when $\textbf{x}$ is far away from the interfacial region. The second term is a
phase field approximation for the line tension energy where a double well potential function
\begin{align}\label{DoubleWell}
\Phi(\eta)=\dfrac{1}{4\xi}(\eta^2-1)^2
\end{align}
is incorporated. $\nabla_{\Gamma}\eta$ is the surface gradient of
$\eta$ which is the projection of $\nabla\eta$ onto the tangent
plane of $\Gamma$. Notice that to make $\nabla_{\Gamma}\eta$ well
defined, the function $\eta$ should be defined away from the
membrane such that $\text{d}\eta/\text{d}\textbf{n}=0$ where
$\textbf{n}$ is the normal vector of $\Gamma$. The enclosed volume
and total area of the membrane are assumed to be invariant.
Meanwhile, the total amount of lipids is conserved. Thus three
constraints are imposed during the minimization of the total energy
(\ref{TotalEnergyPhaseField}):
\begin{align}\label{Constraint}
\int_{\Gamma}\ \text{d}\textbf{x}=\text{A},\quad \int_{\Gamma}\ \text{d}V=\text{Vol},
\quad \int_{\Gamma}\eta(\textbf{x})\ \text{d}\textbf{x}=C.
\end{align}
The constraint $\int_{\Gamma}\eta(\textbf{x})\ \text{d}\textbf{x} = C$ denotes the difference in the surface
areas of the two phases in the sharp interface limit.

\subsection{Axisymmetric Setting}

\begin{figure}
\centerline{\includegraphics[width=3in]{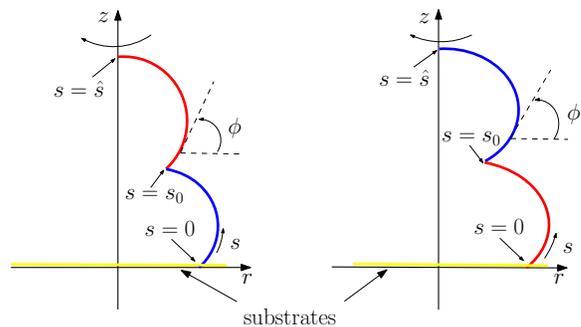}}
\vskip-8pt \caption{Schematic diagrams of axisymmetric two-component adhered vesicle membranes. Blue and red colors indicate the liquid-ordered and the liquid-disordered phases, respectively. $s=s_0$ specifies the phase boundary. Away from the interface, the blue phase has mean curvature bending modulus $c_0+c_1$ and adhesion potential $w(c_0+c_2)P(\text{d}(\textbf{x}))$, while the red phase has mean curvature bending modulus $c_0-c_1$ and adhesion potential
$w(c_0-c_2)P(\text{d}(\textbf{x}))$.}
\label{free} \vskip-10pt
\end{figure}

In the present work, we focus on the axisymmetric membrane adhered on
a flat/curved substrate. In this setting, the membrane surface is determined
by evolving a 2-d curve. A vesicle with a flat substrate is schematically
shown in figure \ref{free}. The axisymmetric vesicle surface is generated by evolving
a curve parameterized by the arc-length $s$, and the total length of the generating
curve is denoted by $\hat{s}$.
The flat substrate is located at $z=0$ with
the two different phases being  distinguished by the
red and blue colors. The transition point from red to blue is located at
$s=s_0$. The figures on the left and right
show different configurations
with the blue phase and the red phase being adjacent to the substrate,
respectively. For easy reference, we denote the left as {\em red-blue}
vesicle membrane, and the right as {\em blue-red} vesicle membrane.
The tangent angle $\phi$ is measured from the radial direction and $r$ is the distance of a point
on the  membrane from the axis of symmetry.

The mean curvature of the vesicle
can be explicitly expressed by $r$ and $\phi$ as
\[
H=\dfrac{1}{2}\Big(\phi'+\dfrac{\sin\phi}{r}\Big),
\]
where prime represents the derivative with respect to arc-length $s$.
The line tension energy term in (\ref{TotalEnergyPhaseField}) becomes
\[
2\pi\sigma\int_0^{\hat{s}}\dfrac{\xi}{2}\eta'^2+\dfrac{1}{4\xi}(\eta^2-1)^2\ \text{d}s.
\]

We nondimensionalize all the parameters and choose $c_0$ to be 1.
Then the phase field model (\ref{TotalEnergyPhaseField}) is reduced to
\begin{align}\label{TotalEnergyAxisymmetry}
\nonumber
E(\eta)=2\pi\int_0^{\hat{s}}&(1+c_1\eta)\big[H-a(1+c_3\eta)\big]^2r\ \text{d}s\\
\nonumber
+2\pi\sigma&\int_0^{\hat{s}}\bigg[\dfrac{\xi}{2}\eta'^2
+\dfrac{1}{4\xi}(\eta^2-1)^2\bigg]r\ \text{d}s\\
&-2\pi\int_0^{\hat{s}}w(1+c_2\eta)
P(\text{d}(\textbf{x}))\ r\ \text{d}s\,.
\end{align}
Additionally, the constraints (\ref{Constraint})
in the axisymmetric case lead to
\begin{align}\label{ConstraintArea}
(\cos T)'=-r,
\end{align}
with $T$ being the arc-length parameter of the reference unit sphere,
\begin{align}\label{ConstraintVol}
\pi\int_0^{\hat{s}}r^2z'ds=\text{Vol},
\end{align}
and
\begin{align}\label{ConstraintLipid}
\int_0^{\hat{s}}\eta(s)rds=\text{C}\,.
\end{align}
The pointwise constraint (\ref{ConstraintArea}), which is in fact equivalent to the one in (\ref{Constraint}), is referred as the lateral incompressibility
condition for the membrane \cite{Das2}.

The shape of the membrane is determined by minimizing the total energy
(\ref{TotalEnergyAxisymmetry}), subject to constraints (\ref{ConstraintArea}),
(\ref{ConstraintVol}) and (\ref{ConstraintLipid}). It satisfies
 the Euler-Lagrange equations given by:
\begin{widetext}
\begin{align}\label{ELEqn}
\nonumber
\big[\widetilde{H}''+\dfrac{r'\widetilde{H}'}{r}+2\widetilde{H}&(H^2-K)+2a\widetilde{H}
H(1+c_3\eta)\big]-2(\mu H +p+\tau \eta H)\\
&+\sigma\xi \eta'^2\phi'
-2\sigma H\Big[\dfrac{\xi}{2}\eta'^2
+\Phi(\eta)\Big]
-w(1+c_2\eta)\bigg(\dfrac{dP}{d(\text{d}(\textbf{x}))}
\dfrac{\delta(\text{d}(\textbf{x}))}{\delta\textbf{n}}-2HP\bigg)=0
\end{align}
\begin{align}\label{EqnEta}
c_1 \big[H-a(1+c_3\eta)\big]^2-2ac_3\widetilde{H}+\sigma\Big[-\xi\Big(\eta''+\dfrac{r'\eta'}{r}\Big)
+\dfrac{d\Phi}{d\eta} \Big]+\tau-c_2wP=0\,,
\end{align}
\end{widetext}
where $\widetilde{H}=(1+c_1\eta)\big[H-a(1+c_3\eta)\big]$ and $\mu, p$ and $\tau$ are the three
Lagrange multipliers for the three constraints, respectively. The other equations from geometry are \cite{Das3, Jenkins}:
\begin{align}\label{OtherEqn}
\phi'=2H-\dfrac{\sin\phi}{r},\quad r'=\cos\phi,\quad z'=\sin\phi.
\end{align}
Boundary conditions are imposed as follows:
\begin{align}\label{BoundaryCondition}
\nonumber
&H'(0)=H'(\hat{s})=0;\ r(0)=r(\hat{s})=0;\\
&\phi(0)=0,\phi(\hat{s})=\pi;\ \eta'(0)=\eta'(\hat{s})=0.
\end{align}

\section{Numerical experiments}
\label{numerical_examples} We numerically solve Eqs (\ref{ELEqn})
through (\ref{OtherEqn}) subject to boundary condition
(\ref{BoundaryCondition}) using the MATLAB ODE solver BVP4C. We
tested the convergence of the computed results in our numerical
simulation.

\subsection{Adhesion with Gaussian potential}
We first present the numerical results of a few adhered vesicles
using the Gaussian adhesion potential.

\begin{figure}[t]
\centerline{\includegraphics[width=3.9in]{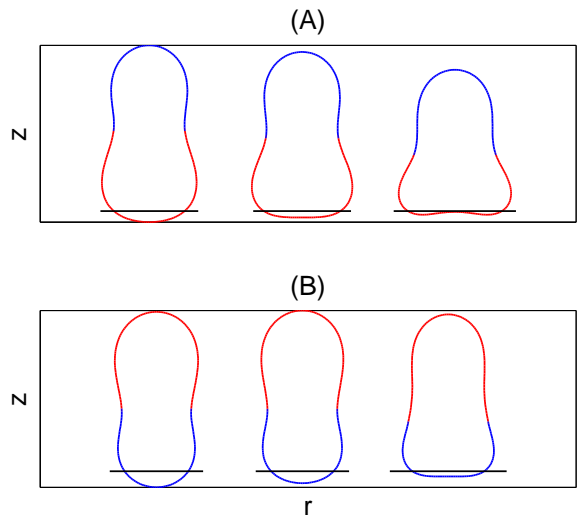}}
\vskip-8pt \caption{Membrane shapes for different $w$. Area=$4\pi$, volume=3.5, area difference
 $C=-0.3$, $\sigma=2$, $c_1=0.4$, $c_3=0$, $a=0$, $\xi=0.01$, $\epsilon=0.15$; (A) $c_2=-0.05$,
  $w=1.0, 1.5, 2.0$, from left to right; (B) $c_2=0.2$, $w=2.0, 3.0, 5.0$, from left to right. }
\label{free_fig1} \vskip-10pt
\end{figure}

In Figure \ref{free_fig1},  several numerical solutions for Eqs.
(\ref{ELEqn}) through (\ref{BoundaryCondition}) are shown. The
curves are the cross-sections of the vesicle shapes with the
blue-colored region representing the liquid-ordered phase and the
red-colored region representing the liquid-disordered phase.
 The
corresponding parameter values used in the experiments are taken as
area=$4\pi$, volume=3.5, $C=-0.3$, $\sigma=2$, $c_1=0.4$, $c_3=0$,
$a=0$, $\xi=0.01$, and $\epsilon=0.15$. Note that $c_1=0.4$ implies
that the ratio
$\lambda_1/\lambda_2=\lambda_{\text{blue}}/\lambda_{\text{red}}$ of
the mean curvature bending moduli between the blue and red phases is
1.4/0.6. Moreover, $a=0$ and $c_3=0$ imply that both phases have
zero spontaneous curvature. Blue-red adhered vesicles on flat
substrate are shown in figure \ref{free_fig1}-A where $c_2=-0.05$,
that is, $w_1/w_2=w_{\text{blue}}/w_{\text{red}}$ in
(\ref{AdhesionPotential}) is equal to 0.95/1.05. Except the
adhesion-associated parameter $w$,
 all the parameters are kept fixed as specified.
Some adhered shapes of red-blue vesicles are shown in figure 2-B for various $w$ and $c_2=0.6$.
Notice that there is a slight protrusion of vesicles into the flat substrate.
This is due to the lack of repulsive effect in the Gaussian form of the adhesion
potential.

\begin{figure}[t]
\centerline{\includegraphics[width=3.9in]{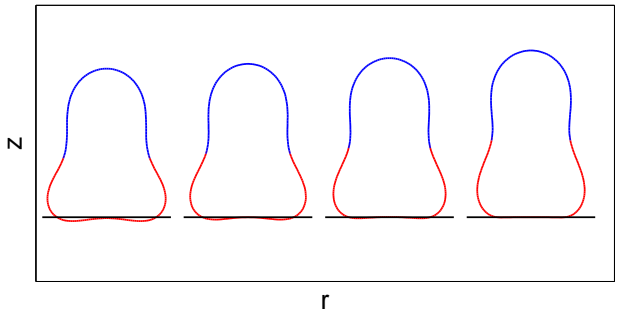}}
\vskip-8pt \caption{Convergence of adhered membranes as $\epsilon\rightarrow 0$.
 $\epsilon=0.15, 0.10, 0.05, 0.01$, from left to right. Area=$4\pi$, volume=3.5,
  area difference $C=-0.3$, $\sigma=2$, $c_1=0.4$, $c_2=-0.05$, $c_3=0$, $a=0$,
  $\xi=0.01$, $w=2$.}
\label{free_fig2} \vskip-10pt
\end{figure}

\begin{figure}[b]
\centerline{\includegraphics[width=3.7in]{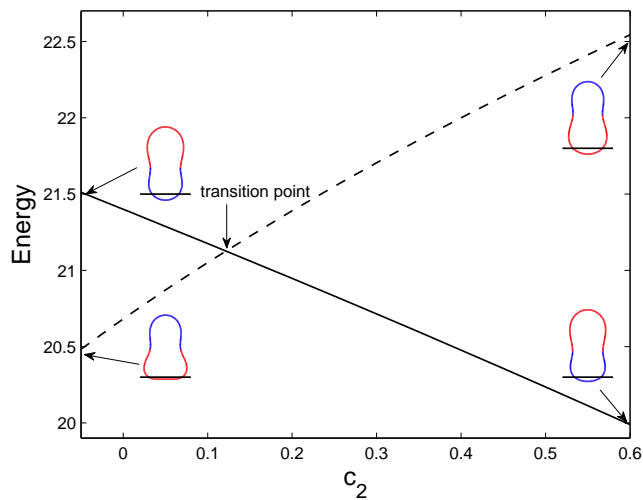}}
\vskip-8pt \caption{Energy comparison between blue-red vesicle and red-blue one.
$w=2, c_1=0.3$ are fixed. $c_2$ varies from -0.05 to 0.6.
The solid curve represents energy versus $c_2$ for the red-blue vesicle, while
the dash curve is for the
blue-red vesicle. The shapes of the vesicles at the four end-points of the curves
 are shown. Area=$4\pi$, volume=3.5, area difference $C=-0.3$, $\sigma=2$,
 $c_3=0$, $a=0$, $\xi=0.01$, $\epsilon=0.15$.}
\label{free_fig3} \vskip-10pt
\end{figure}

Convergence of the adhered vesicle shapes as $\epsilon$ approaches
zero is presented in figure \ref{free_fig2}. Theoretically, by the
standard asymptotic analysis, $\eta$ tends to converge to
$\tanh\big(\frac{s-s_0}{\sqrt{2}\xi}\big)$ when $\xi$ approaches zero, where
$s_0$ indicates the location of the interface which depends on the
area difference $C$ (see appendix A). Numerically, for relatively
larger $\epsilon$, the shapes of two-component vesicle membranes
protrude into the flat substrate more significantly.
 As $\epsilon$ gets smaller, the protrusion becomes less visible
and finally vanishes in the limit of $\epsilon\to 0$, that
is the case corresponding
to the effective contact potential \cite{Das1,JZhang,Lipowsky2}
\begin{align}
W(\textbf{x})=
 \left\{\begin{array}{l} w\quad d(\textbf{x})=0 \\ 0\quad \text{otherwise}. \end{array}\right.
\end{align}

Figure \ref{free_fig3} shows the energy comparison between blue-red
vesicle membrane and red-blue one under the influence of adhesion.
Here area=$4\pi$, volume=3.5, $C=-0.3$, $\sigma=2$, $c_1=0.3$,
$c_3=0$, $a=0$, $\xi=0.01$, $\epsilon=0.15$, $w=2.0$ are fixed. The two
shapes on the left correspond to parameter $c_2=-0.05$ with
$w_{\text{blue}}/w_{\text{red}}=0.95/1.05$. The blue-red vesicle,
which is more deformed from the free vesicle shape, has lower energy
and is more stable than the red-blue one in this case. On the other
hand, the two shapes on the right correspond to $c_2=0.6$ leading to
$w_{\text{blue}}/w_{\text{red}}=1.6/0.4$. There, the red-blue
vesicle which is deformed more from the free shape has lower energy
and is more stable than the blue-red one. In general, for $c_2 < 0$,
$w_{\text{red}}$ is larger than $w_{\text{blue}}$ and blue-red
membrane is more deformed from a free shape. Whereas, for $c_2 > 0$,
$w_{\text{blue}}$ is larger than $w_{\text{red}}$ and red-blue
membrane is more deformed from a free shape.

\begin{figure}[t]
\centerline{\includegraphics[width=3.9in]{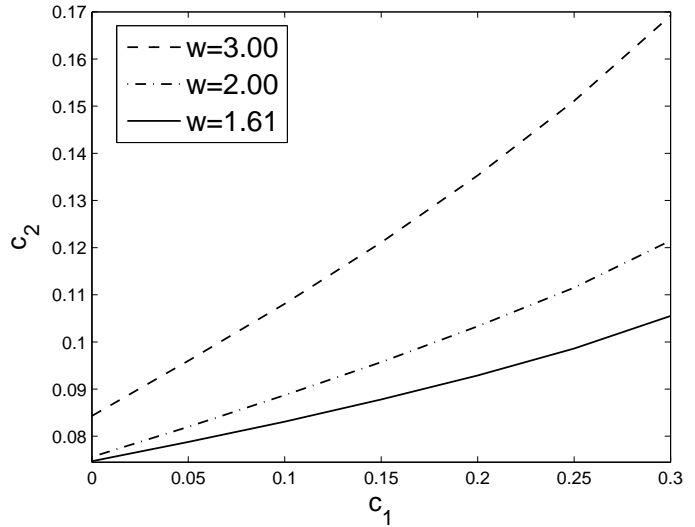}}
\vskip-8pt \caption{Transition curves for w=1.5, 2.0, 3.0. Above each curve, red-blue membrane is more stable;
the blue-red membrane is more stable below the curves. The other parameter values are: Area=4$\pi$, volume=3.5,
area difference $C=-0.3$, $\sigma=2$, $c_3=0$, $a=0$, $\xi=0.01$, $\epsilon=0.15$.}
\label{free_fig4} \vskip-10pt
\end{figure}

The energy comparison of the four shapes as discussed above may lead
us to believe that the vesicle whose component adjacent to the
substrate endures stronger adhesion is more stable. However, this is
not always the case. As seen in figure \ref{free_fig3}, there is an
anomalous region with $c_2$ between zero and 0.1217, and the blue
phases suffer stronger adhesion but the blue-red vesicle is more
stable. Outside this region, the shape with stronger adhesion on the
component adjacent to the substrate is more stable. We observe that
the existence of the region (0, 0.1217) is due to the nonzero values
of $c_1$ and $C$ which model, in the two-component system, the
effects due to the differences in the bending moduli and the surface
areas of the two phases. When both $c_1$ and $C$ approach zero which
is the limit where both components have the same bending moduli
and equal surface areas,  the multi-component vesicle reduces into
a single component vesicle. And the anomalous region shrinks to the point
$c_2=0$, also the transition point converges to $c_2=0$,

In addition, the transition point in figure \ref{free_fig3} strongly depends on $c_1$ and the adhesion potential $w$. The dependence of the transition points on $c_1$ is shown in figure \ref{free_fig4}. For various $w=3.00, 2.00, 1.61$, we  plot the transition curve $c_1$ versus $c_2$. The red-blue vesicle, corresponding to the parameter pair $(c_1, c_2)$ above each curve, is more stable; while the blue-red vesicle is more stable if the parameters $c_1, c_2$ are chosen from the region below the transition curves.

\begin{figure}[t]
\centerline{\includegraphics[width=3.9in]{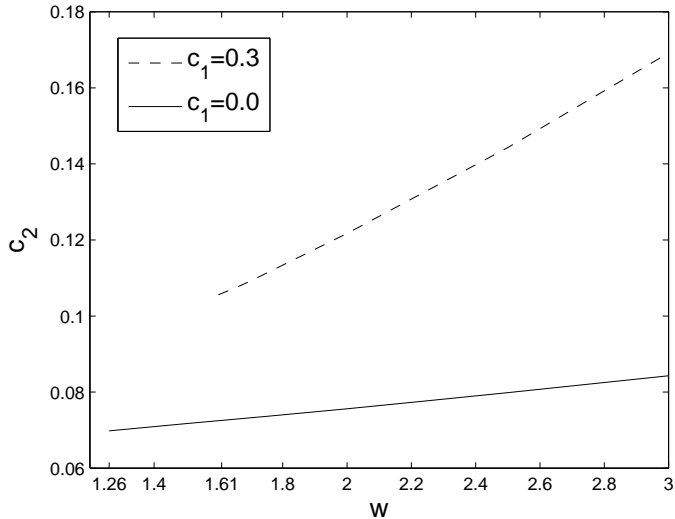}}
\vskip-8pt \caption{Dependence of $c_2$ on $w$ for phase transition when $c_1=0.3$ and 0.0, respectively. Area=4$\pi$, volume=3.5, area difference $C=-0.3$, $\sigma=2$, $c_3=0$, $a=0$, $\xi=0.01$, $\epsilon=0.15$.}
\label{free_fig5} \vskip-10pt
\end{figure}

Figure \ref{free_fig4} shows that the transition always occurs for
$c_2>0$. One may wonder if this is universally  true for  any
adhesion potential $w$. The answer is provided via figure
\ref{free_fig5}, the transition curve $w$ v.s. $c_2$ for fixed
$c_1$. In figure \ref{free_fig5}, the dash curve corresponds to
$c_1=0.3$, and the solid curve corresponds to $c_1=0$. Then we
obtain the dependence of $c_2$ on $w$ when the transition occurs.
For $c_1=0.3$, $w$ varies from 1.61 to 3; while for $c_1=0.0$, $w$
varies from 1.26 to 3. Above the transition curve, the red-blue
vesicle is the more stable one; while below the curve, the blue-red
vesicle is more stable. The bound blue-red vesicle with given
parameters will break away from the adhered state around $w=1.60
(1.25)$ when $c_1=0.3 (0.0)$ and change to an unbound (free)
vesicle. We can thus claim that $c_2$ is always positive for any
bound adhesion potential $w$ when $c_1$ is fixed.

\begin{figure}[t]
\centerline{\includegraphics[width=3.9in]{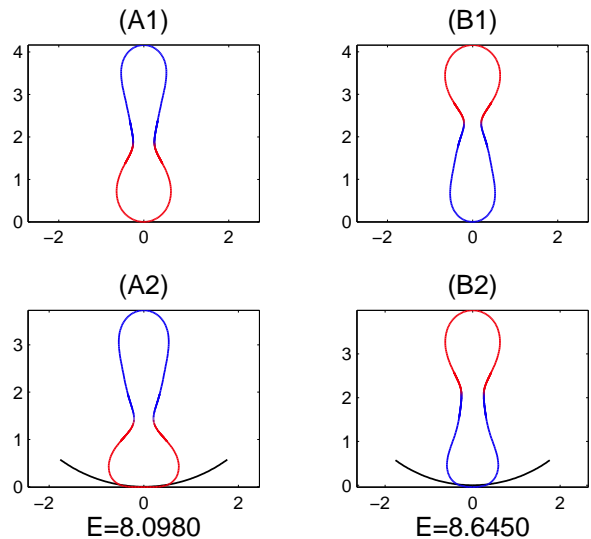}}
\vskip-8pt \caption{Effect of spontaneous curvatures on the vesicle shapes. Representative shapes of vesicle membranes are shown.  Area=4$\pi$, volume=2.7, area difference $C=0$, $\sigma=1$, $c_1=0.0$, $c_2=0$ $c_3=0.5$, $a=2/3$, $\xi=0.01$, $\epsilon=0.05$, $w=2$.}
\label{free_fig8} \vskip-10pt
\end{figure}

\begin{figure}[b]
\centerline{\includegraphics[width=3.9in]{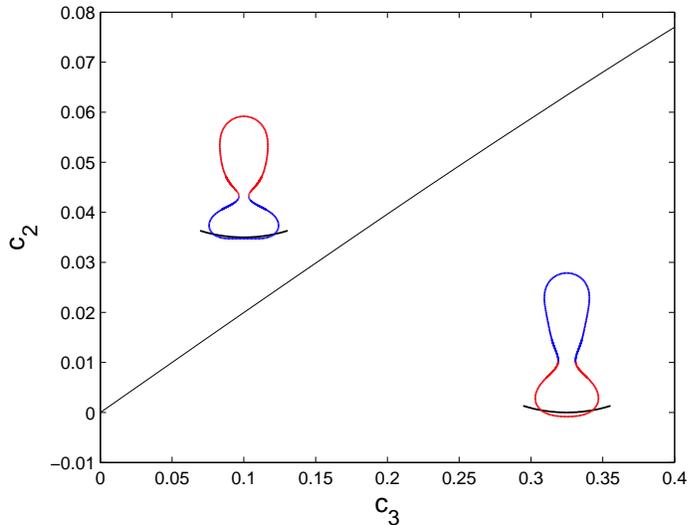}}
\vskip-8pt \caption{Dependence of $c_2$ on $c_3$ for phase transition. Representative shapes of vesicle membranes are shown above or below the transition curve. area=$4\pi$, volume=$2.7$, area difference $C=0$, $\sigma=1$, $c_1=0.0$, $a=2/3$, $\xi=0.01$, $\epsilon=0.15$, $w=2$.}
\label{free_fig9} \vskip-10pt
\end{figure}

We also take the effects of the spontaneous curvature and the
substrate curvature into consideration \cite{Ursell,Das4}. In figure
\ref{free_fig8} the flat substrate is now replaced by a concave-up
spherical substrate with radius $R=3$. Here area=$4\pi$,
volume=$2.7$, area difference $C=0$, $\sigma=1$, $c_1=0$, $c_2=0$,
$c_3=0.5$, $a=2/3$, $\xi=0.01$ and $\epsilon=0.05$. We take $w=0$
and $w=2$ respectively to find both free vesicles and adhered
vesicles. Differing from the previous figures in this subsection,
the red and blue colors in this experiment indicate the phases
having different spontaneous curvatures. The blue phase has a bigger
spontaneous curvature $a(1+c_3)=1$ while the red phase has a smaller
one $a(1-c_3)=1/3$. A1 and B1 show the free vesicles with specified
parameters. Notice that the red phase, which possesses a smaller
spontaneous curvature, has shapes less elongated than the blue phase.
A2 and B2 are adhered blue-red and red-blue vesicles. By comparing
the total energy of these two adhered vesicle, we find out that the
blue-red vesicle with the red phase at bottom, whose spontaneous
curvature matches with the curvature of the curved substrate, is
more stable than the red-blue vesicle.

We choose here a volume with value 2.7 which is smaller than the
previously chosen value of 3.5 because larger osmotic pressure
difference makes the effect of spontaneous curvatures on the vesicle
shapes more visible.

Similar to the discussion of stability transition in figure \ref{free_fig4} and figure \ref{free_fig5}, when the spontaneous curvature effect is taken into account, we can also find the dependence of transition points on $c_1, c_2, c_3, w$. Without repeatedly showing many figures, we present here only the transition curve $c_3$ v.s. $c_2$ in figure \ref{free_fig9} with $c_1=0, w=2$ fixed. Other parameters are set as area=$4\pi$, volume=$2.7$, area difference $C=0$, $\sigma=1$, $a=2/3$, $\xi=0.01$, $\epsilon=0.15$. Above the transition curve, a representative red-blue vesicle is shown which is more stable than the blue-red vesicle; while below the transition curve, the blue-red vesicle is more stable.

\subsection{Adhesion with Leonard-Jones potential}

When the Gaussian adhesion potential is used, as  seen in the above,
slight protrusion of vesicles into the substrate may occur. If this
protrusion is of any practical concern, one possible way to avoid the
protrusion is to replace the Gaussian adhesion potential by a Leonard-Jones
type potential. We demonstrate, by employing the following potential,
\begin{align}\label{LJ}
P(\text d(\textbf{x}))=-4\bigg[\bigg(\dfrac{\beta}{\text{d}(\textbf{x})}\bigg)^{\alpha}
-\bigg(\dfrac{\beta}{\text{d}(\textbf{x})}\bigg)^{\alpha/2}\bigg]\,,
\end{align}
that the shapes of adhered vesicles without any protrusion can be
obtained using our phase field formulation. The key difference
between the Gaussian type potential (\ref{AdhesionPotential}) and
the Leonard-Jones type potential (\ref{LJ}) is that Gaussian
potential is globally attractive, while there is a narrow region
with $\text{d}(\textbf{x})$ between zero and $\beta$ where the
Leonard-Jones type potential is repulsive. Such a repulsive region
can prevent the vesicle membrane from protruding into the substrate.

\begin{figure}[t]
\centerline{\includegraphics[width=3.9in]{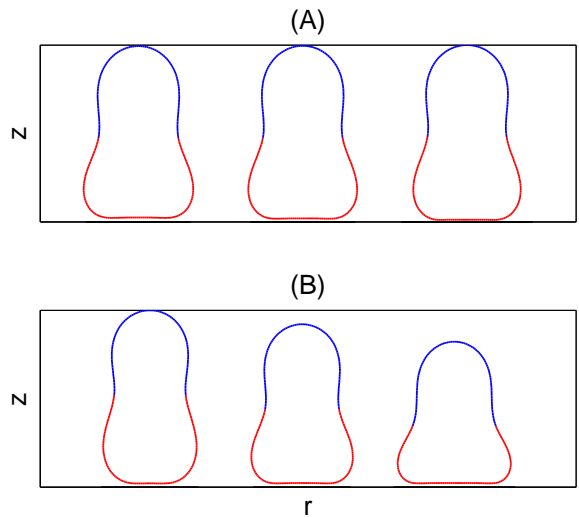}}
\vskip-8pt \caption{Membrane shapes for Leonard-Jones type potential. (A) $w=2$, $\beta=0.05, 0.035, 0.025$,
from left to right; (B) $\beta=0.05$, $w=1.2, 2.0, 2.8$, from left to right. Other parameters:
Area=$4\pi$, volume=3.5, area difference $C=-0.3$, $\sigma=2$, $c_1=0.3$, $c_2=-0.05$,
$c_3=0$, $a=0$, $\xi=0.01$, $\alpha=5$ }
\label{free_fig7} \vskip-10pt
\end{figure}

An axisymmetric simulation is shown in figure \ref{free_fig7} where
we take $\alpha=5$. In figure \ref{free_fig7}(A), $w=2$ is fixed,
and the parameter $\beta$ is decreased from left to right. We can
see that the repulsive region between substrate and vesicle membrane
narrows down. The shape deformation under the influence of adhesion
is shown in figure \ref{free_fig7}(B) where the repulsive region is
fixed but $w$ is increased from left to right. The stability
transition curves, similar to that obtained previously for Gaussian
potential, can also be obtained for Leonard-Jones type potential.
Given the similarities in the findings, we do not repeat the
discussion here.

\subsection{Promotion of phase separation}

In this subsection, two numerical experiments are presented to
support Gordon and coworkers' experimental observation that adhesion
may promote the phase separation  \cite{Gordon}. The experiments
involve the phase field simulations for relatively large diffuse
interface parameter $\xi$ which is consistent to the experimental
setting. The Leonard-Jones type adhesion potential is employed in
this subsection.

\begin{figure}[b]
\centerline{\includegraphics[width=3.7in]{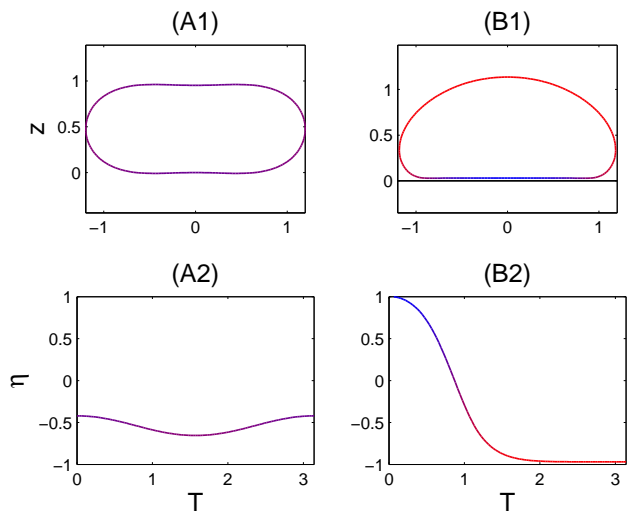}}
\vskip-8pt \caption{Adhesion induces phase separation. An almost homogeneous free vesicle is phase-separated
by a Leonard-Jones adhesion.}
\label{Promotion02} \vskip-10pt
\end{figure}

In the first experiment, the function $\eta(s)$ is viewed as a
chemical composition function, which can be considered as a
compositional fluctuation around a homogeneous state with
composition $\eta_0$.  The constraint (\ref{ConstraintLipid}) is
specified as
\begin{align}\label{ConstraintComposition}
\int_0^{\hat{s}}\eta(s)rds=\int_0^{\hat{s}}\eta_0rds
\end{align}

We now demonstrate that for an equilibrium free vesicle with
associated $\eta(s)$ almost homogeneous, namely $\eta(s)\approx
\eta_0$, after adding the adhesion, it displays  the phase
separation behavior with $\eta(s)$ changing into a tanh-like profile
representing two distinct phases.

\begin{figure}[b]
\centerline{\includegraphics[width=3.9in]{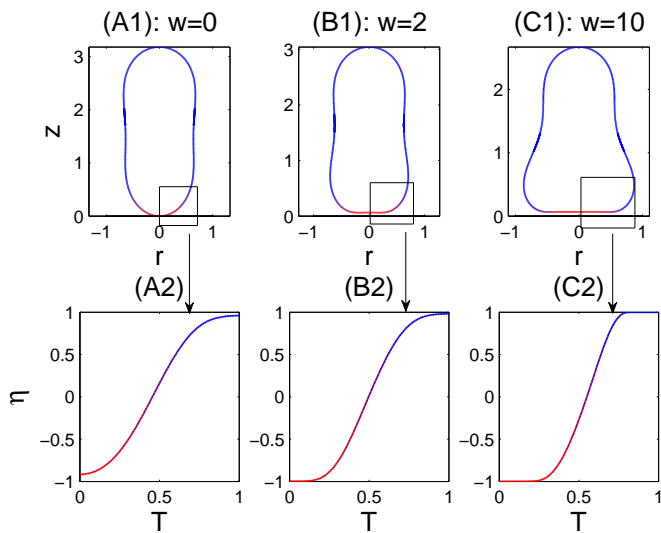}}
\vskip-8pt \caption{Promotion of phase separation. As $w$ gets bigger, the two phases of $\eta$ near the
interfacial region are separated more. Area=4$\pi$, volume=3.5, area difference $C=1.7$,
$\xi=0.13$, $\sigma=1$, $c_1=0.05$, $c_2=-1.00$, $c_3=0$, $\zeta=1$, $\beta=0.05$, $\alpha=5$.}
\label{free_fig6} \vskip-10pt
\end{figure}

The numerical experiment is presented in figure \ref{Promotion02}.
In (A1-2), for $\text{area}=4\pi, \text{volume}=3.5,
\eta_0=-1/\sqrt{3}, \xi=0.33, \sigma=5/(1-0.8\eta_0),
c_1=0.8/(1-0.8\eta_0), c_2=1.0/(1-1.0\eta_0), c_3=0$, a free vesicle
is computed first. The chemical composition function $\eta$ and the
associated vesicle profile shown in (A1-2) correspond to the stable
energy minimum (in the absence of adhesion). The change in
composition is relatively small, representing a state of mixed
phases in much of the vesicle.  By adding Leonard-Jones adhesion
with $w=5(1-1.0\eta_0)/(1-0.8\eta_0), \beta=0.023, \alpha=5$, an
adhered vesicle with associated $\eta$ corresponding to the new
energy minimizer is shown in (B1-2). A phase separation occurs with
a much more significant phase difference
$\max(\eta)-\min(\eta)=1.9703$ in (B2). In comparison, we have
$\max(\eta)-\min(\eta)=0.2345$ in (A2), which is only $11.90\%$ of
the phase difference for adhered vesicle. We thus see that adhesion
can significantly promote phase separation. Furthermore, the bending
modulus of the phase-mixed vesicle $\lambda_{\text{mixed}}$ is
roughly around $1/(1-0.8\eta_0)$, and the bending moduli of the
phase-separated vesicle are $\lambda_{\text{blue}}=
2.2649/(1-0.8\eta_0)$ and
$\lambda_{\text{red}}=0.6873/(1-0.8\eta_0)$. These parameter values
imply that $\lambda_{\text{red}}\approx \lambda_{\text{mixed}}$ and
$\lambda_{\text{blue}}/\lambda_{\text{red}}=3.2954$ which agree with
the ratios of the bending stiffnesses of the ordered, the disordered
and the mixed phases available in the literature
\cite{Gordon,Roux,Baumgart2,Baumgart1}.

The effect of adhesion on phase separation can be further
demonstrated in the next experiment. Here again, we take the order
parameter $\eta(s)$ to be a labeling function for relative lipid
density which stays within $[-1,1]$. To model a wider diffuse
interfacial layer corresponding to a relatively large $\xi$ and
yet maintain the bound on $\eta$, we replace the double well
potential function (\ref{DoubleWell}) by the
following double obstacle potential function \cite{elliott}:
\begin{equation}\label{DoubleObstacle}
(1+\eta)\ln(1+\eta)+(1-\eta)\ln(1-\eta)
+\gamma(1-\eta^2)-2\ln 2\;.
\end{equation}
where $\gamma=1+2\ln2$ is a constant describing the height of
the potential barrier.

Figure \ref{free_fig6} shows a numerical experiment while we choose
$\xi=0.13$ so that for unbounded vesicles, there would be a wide
interfacial regions with less dramatic phase separation effect. Then
for various $w=0, 2, 10$, we compute the corresponding adhered
vesicle shapes and the corresponding lipid density functions $\eta$.
We highlight the densities near the interfacial region (the boxed
portion of the vesicle profile). With the plots using the same
scaling in the arc-length, one can see that with a larger adhesion
strength $w$, the interfacial layer gets narrower and the separation
between the red and blue phases becomes sharper and more dramatic.

\section{Conclusion}
In this work, we develop a phase field model for the adhesion of the multi-component vesicle membrane to a flat/curved substrate. Some representative vesicle shapes are presented. The influence of $c_2$, which measures the contrast of the adhesion
effect for the different phases, on the stability of membrane shapes is discussed
here. It turns out that the multi-component vesicle, whose lower part suffers stronger adhesion, is more stable in most cases. We also consider the the phase transition from blue-red vesicle to red-blue vesicle, and the influence of other parameters such as
the relative contribution of the adhesion and the bending energy. The effect of spontaneous curvature is also numerically observed by examining the adhesion of the vesicles on a
curved substrate. We also present vesicle shapes corresponding to the Leonard-Jones type adhesion potential to show that the repulsive region can effectively prevent the
protrusion of the vesicles into the substrates should this be of practical
concern. Finally, we numerically examine the fact that adhesion can promote the phase separation for multi-component vesicle membrane.

Although the numerical studies presented here are focusing on the axisymmetric configuration, the phase field formulation is applicable to more general
settings such as in the study of the interaction between the multi-component
vesicle and a patterned substrate. It can also be used to extend the study of
 membrane-mediated particle interactions \cite{deserno2003, deserno2004}
to  multi-component vesicles. The present phase field formulation of the multi-component vesicle and substrate interaction can also be extended in several directions. For instance, by incorporating the Gaussian curvature contributions to the bending energy, fusion of multi-component vesicles can be studied. By adding an entropic contribution to the free energy, we can also consider the weak adhesion
regime where fluctuation of the vesicle shape due to thermal excitation
plays an important role.

\appendix

\section{Asymptotic analysis for $\eta$}
In section \ref{numerical_examples},  the convergent behavior of the
adhered vesicles as the interfacial width $\epsilon$ approaches zero
is briefly mentioned. It is claimed that the phase field function $\eta$
approaches $\tanh\big(\frac{s-s_0}{\sqrt{2}\xi}\big)$ as the parameter $\xi$
approaches zero. A boundary layer calculation is carried out to support
such an observation\cite{Holmes}.

The lipid density function $\eta(s)$ is governed by the Eq (\ref{EqnEta}) and can be rewritten as
\begin{align}\label{EqnEtaRewrite}
\nonumber
\xi\Big(c_1 \big[H-a(1&+c_3\eta)\big]^2-2ac_3\widetilde{H}+\tau-c_2wP\Big)r\\
&+\sigma\Big[-\xi^2(\eta'r)'
+(\eta^2-1)\eta r\Big]=0
\end{align}

An asymptotic analysis in the sharp interface limit is carried out
as follows. For simplicity, we only consider the O(1) terms for
outer and inner layers here. The inner layer and the outer layer are
the regions around the interface and away from the interface,
respectively.

Within the outer layer, $\eta$ does not change much with respect to
the arc-length and we expand $\eta(s)$ as
\begin{align}\label{OuterExpansionO1}
\eta(s)\sim \eta_0(s)+\xi\eta_1(s)+\ldots\,,
\end{align}
We substitute (\ref{OuterExpansionO1}) into (\ref{EqnEtaRewrite}) and obtain
an equation for $\eta_0$ at the lowest order:
\[
(\eta_0^2-1)\eta_0r=0\,.
\]
The solutions are
\[
\eta_0=0,\pm 1\,,
\]
and we choose $\eta = 1$ and $\eta = -1$ in the two sides of the interface
to represent the blue and red phases, respectively. Here let us assume
$\eta=-1$ in the side $s<s_0$; and $\eta=1$ in the side $s>s_0$.

In the inner layer, we expect $\eta$ to vary rapidly from $-1$ to $1$ and
introduce a new stretched arc-length variable
\[
S=\dfrac{s-s_0}{\xi^{\mu}}\,,
\]
where $\mu$ is an yet undetermined parameter. We now express $\eta$ as a
function of the stretched variable $S$ and expand $\eta$ as
\begin{align}\label{InnerExpansionO1}
\widetilde{\eta}(S)\sim \widetilde{\eta}_0(S)+\xi\widetilde{\eta}_1(S)+\cdots\,.
\end{align}
Notice that
\[
r(s)=r(s_0+\xi^{\mu} S)=r(s_0)+\xi^{\mu} Sr'(s_0)+O(\xi^{2\mu}),
\]
Upon rewriting Eq. (\ref{EqnEtaRewrite}) in terms of the stretched
variable $S$ and, subsequently, using Eq.\ (\ref{InnerExpansionO1})
we obtain
\begin{align}\label{EquationInner}
\nonumber
-&\xi^{2-2\mu}r(s_0)\dfrac{d^2\widetilde{\eta}_0}{dS^2}
-\xi^{2-\mu}r'(s_0)\dfrac{d}{dS}\Big(S\dfrac{d\widetilde{\eta}_0}{dS}\Big)\\
&+(\widetilde{\eta}_0^2-1)\widetilde{\eta}_0\big[r(s_0)+\xi^{\mu} Sr'(s_0)\big]+O(\xi^{\mu})=0
\end{align}
There are two possible choices for $\mu$. If $\mu$ balances the second
and third terms, namely, $2-\mu=0$. Then the leading order term is:
\[
-r(s_0)\dfrac{d^2\widetilde{\eta}_0}{dS^2}=0
\]
which implies $\widetilde{\eta}_0=aS+b$. However,
this solution does not satisfy the matching conditions
\[
\widetilde{\eta}_0(-\infty)=\eta_0(0) = -1,\quad \widetilde{\eta}_0(+\infty)=\eta_0(\hat{s}) = 1\,.
\]

Another choice is to balance the first and third terms of equation
(\ref{EquationInner}), namely, $\mu=1$, and one gets the leading
order term:
\[
-r(s_0)\dfrac{d^2\widetilde{\eta}_0}{dS^2}+(\widetilde{\eta}_0^2-1)\widetilde{\eta}_0r(s_0)=0.
\]
If the boundary condition $\widetilde{\eta}_0(-\infty)=-1, \widetilde{\eta}_0(+\infty)=+1$ are imposed,
a typical solution of the above nonlinear equation, which satisfies
the matching condition, is
\[
\widetilde{\eta}_0(S)=\tanh\Big(\dfrac{S}{\sqrt{2}}\Big).
\]

Then the composite solution of (\ref{EqnEtaRewrite}), given by
\begin{center}
inner solution + outer solution - matching solution,
\end{center}
takes the form
\[
\widetilde{\eta}_0(S)+\eta_0^{-}(s)-\eta_0^{-}(0)
+\eta_0^{\text{+}}(s)-\eta_0^{\text{+}}(\hat{s}),
\]
and explicitly
\[
\eta(s)\sim \tanh\Big(\dfrac{s-s_0}{\sqrt{2}\xi}\Big)+\ldots
\]


\bibliography{mad}

\end{document}